\pdfoutput=1
\documentclass[aps,pre, groupedaddress,twocolumn]{revtex4-2}
\usepackage{lipsum}
\usepackage{float}
\usepackage{mathtools}
\usepackage{graphicx}
\usepackage{subfigure}

\usepackage{dcolumn}
\usepackage{amsmath}    
\usepackage{amssymb}
\usepackage{bm}
\usepackage{latexsym}
\usepackage{verbatim}
\usepackage[normalem]{ulem}
\usepackage{color}
\usepackage{parskip}
\usepackage{lipsum}
\usepackage{setspace}
\usepackage{hyperref}
\usepackage{cleveref}
\def\beq{\begin{equation}}
\def\eeq{\end{equation}}

\def\beq{\begin{equation}}                          \def\eeq{\end{equation}}                          
\def\bea{\begin{eqnarray}}                          \def\eea{\end{eqnarray}}

\DeclareRobustCommand{\uvec}[1]{{%
  \ifcsname uvec#1\endcsname
     \csname uvec#1\endcsname
   \else
    \bm{\hat{\mathbf{#1}}}%
   \fi
}}

\preprint{}
\begin{document}
\preprint{}
\title{Synchronized Rotations 
of Active Particles on Chemical Substrates}
\author{Pathma Eswaran$^{1}$}
\email{pathma.eswaran.phy19@iitbhu.ac.in}
\author{Shradha Mishra$^{1}$}
\email{smishra.phy@iitbhu.ac.in}
\affiliation{$^{1}$Department of Physics, Indian Institute of Technology (BHU), Varanasi, India 221005}

\begin{abstract}
Many microorganisms use chemical `signaling' - a quintessential self-organizing strategy in non-equilibrium - that can induce spontaneous aggregation and coordination in behavior. Using synthetic signaling as a design principle, we construct a minimal model of active Brownian particles (ABPs) having soft repulsive interactions on a chemically quenched patterned substrate. The interplay between chemo-phoretic interactions and activity is numerically investigated for a proposed variant of the Keller-Segel model for chemotaxis. Such competition not only results in a chemo-motility-induced phase-separated state but also a new cohesive clustering phase with synchronized rotations. Our results suggest that rotational order can emerge in systems by virtue of activity and repulsive interactions alone without an explicit alignment interaction. These rotations can also be exploited by designing mechanical devices that can generate reorienting torques using active particles.
\end{abstract}
\maketitle

\section{Introduction}
Active matter refers to any collection of entities that individually and dissipatively break time-reversal symmetry and are innately out of equilibrium \cite{rmpbechinger2016, annalsofphys2005, vicsek2012, rmp2013}. The living world is overwhelmingly constituted by active matter in the form of cells \cite{cell}, flocks of birds \cite{physicstoday2006}, human crowds \cite{humancrowd1, humancrowd2}, etc. Active units not only possess exciting features as a collection but also show intriguing individual dynamics and reach a statistical steady state in response to an external stimulus 
that is central to many fascinating behaviors in active systems, viz. collective foraging \cite{foraging}, swarming of bacteria \cite{bacteria1, bacteria2, bacteria3}, dynamical clustering in active colloids \cite{activecolloids}, etc. Several of these collective effects stem from alignment mechanisms.

The response of agents to a stimulus - customarily modeled by field variations in density \cite{densityfield}, chemical potential \cite{chemicalfield}, polarization \cite{polfield, pol2} - has finite effects on the spatiotemporal self-propulsion speeds of the agents that often lead to long-range anisotropic interactions. The effect of quenched (time-independent) disorder or stimulus in the dynamical phases of self-propelled particles \cite{coldperuani, vivekepje, patternedBech, quench} is a topic of great interest but is lacking in its representation in literature. With the rapid development of synthetic microswimmers, it has become easier to employ synthetic signaling as a design principle to create and study pattern formations \cite{acc1, acc2,activerotors}. For example, the response of active agents to a chemo-phoretic field and its effect on the non-equilibrium phenomena unique to active systems, motility-induced phase separation (MIPS), \cite{arXiv} and chemotactic stabilization of hydrodynamic instabilities in active suspensions \cite{instability} has been studied. Furthermore, the interplay between steric, chemo-phoretic interactions and activity, leads to the emergence of a phase-separated state very recently coined as the chemo-motility-induced phase-separated (CMIPS) state \cite{softmatter2023}. 

On the other hand, the effect of surface interactions and morphology on motility can be riveting \cite{bact_morph}. The motion of a Brownian particle as it flows through periodically modulated potential-energy landscapes in two dimensions experiences a crossover from free-flowing to locked-in transport that depends on the periodicity of the landscape \cite{morphology}. A self-propelled colloid faces a competition between hindered diffusion from the trapping potential on a periodic crystalline surface and enhanced diffusion due to active motion \cite{periodicsurface}. Further, a periodic arrangement of obstacles on the substrate is found to enhance the persistent motion of an ABP and induce directionality in its motion \cite{sudipta}.

Such studies motivated us to pursue a quench disorder framework for a collection of ABPs and to look for changes in pattern formations and motility. In this work, we achieve the same by exposing the well-studied collective ABP problem \cite{softmatterpritha2018, prl2012filly} to the Keller-Segel \cite{KS1}-\cite{KS2} model of chemotaxis (swimming up chemical gradients). The presence of a quenched chemical disorder introduces an induced phase-separated state at finite motility, known as CMIPS. The interplay between chemo-phoretic interactions and activity suppresses the dynamical phases that a quench-free ABP problem would otherwise produce.  In addition to CMIPS, we obtain a non-trivial dynamical phase of clusters with synchronized rotations. The emergence of such a phase is accompanied by a cooperative balance between the active force and the chemical force. The rotational sense is picked up by the clusters naturally and is analogous to the spontaneous breaking of symmetry in other condensed matter systems. 

The remainder of the article is organized as follows. In section II, we discuss the model for chemotaxis and numerical details for the Brownian simulations. In section III, we present the single-particle model and the interacting model. The state diagram as a function of activity and steepness of the chemical gradient, the steady-state structural behavior, the dynamical characteristics of the phases, and the phase transitions are described for the latter case. We summarize our significant findings in section IV and suggest directions for future work.

\section{Model and Numerical details}
A collection of ABPs with radii $a_0$ having a self-propulsion speed $v_0$ are simulated on a two-dimensional surface with a patterned chemical concentration. The steric force $\textbf{F}_{ij}$ between two disks $\textit{i}$ and $\textit{j}$ is short-ranged and repulsive: $\textbf{F}_{ij}=-k(2a_{0}-r_{ij})\hat{\textbf{r}}_{ij}$, if $r_{ij} < 2a_{0}$ and $\textbf{F}_{ij}=\textbf{0}$ otherwise. Here, $\textbf{r}_{ij} = \textbf{r}_j - \textbf{r}_i$. In addition to the steric repulsion, particles also experience a time-invariant periodic chemical concentration on the substrate:  $c(\textbf{r}) = h_0  sin(\frac{2\pi x}{\lambda}) sin(\frac{2\pi y}{\lambda})$ with wavelength $\lambda$ and amplitude $h_0$. Within a distance of one wavelength, both a hill and a valley of the chemical potential occur in both x and y directions. Each local minima of the patterned $c({\bf r})$ can be treated as a separate subsystem containing a sufficient number of ABPs. Activity and phoresis are decoupled, making these particles different from ones that are inherently chemically active \cite{chemact1, chemact2}. Then, the motion of an ABP self-propelling in a direction \textbf{p}$_{i}$(t) = $(cos\theta_{i}(t),sin\theta_{i}(t))$ is given by the following over-damped equations :
\begin{equation}
    \partial_t \textbf{r}_i = v_0 \textbf{p}_i(t) + \beta_D \nabla c(\textbf{r}_{i},t) + \mu \sum_{j\neq i} \textbf{F}_{ij}\label{eq3} 
\end{equation}
\begin{equation}
    \partial_t \textbf{$\theta$}_i = \beta \textbf{p}_i(t)\times \nabla c(\textbf{r}_{i},t) + \eta^R_i(t)\label{eq4} 
\end{equation}

Equations \ref{eq3} and \ref{eq4} model the response of active particles to the local chemical gradient drawing from the Keller-Segel model of chemotaxis. \textcolor{black}{$\beta_D$ is the chemophoretic coupling coefficient measuring the translational diffusion in response to the force exerted by the chemical field onto the center of mass of the particle. $\beta$ is the chemotactic coupling coefficient measuring angular diffusion arising from the reorienting torques exerted by the local gradients. The strength of orientational fluctuations in the system is governed by the Gaussian white noise $\eta_i^R$ of strength $D_{R}$ with zero mean and delta correlation, whereas $\beta$ is to be viewed as the true chemotactic response.} The swimming direction of the particle is chemo-attractive if $\beta_D, \beta>0$ (motion towards the chemical gradient) and chemo-repulsive if $\beta_D, \beta<0$ (motion away from the chemical gradient) for the position and orientation respectively. $\beta$ and $\beta_D$ are in principle independent and can have different signs and magnitudes. For example, a passive
particle can undergo phoresis but not taxis. In this study, we fix $\beta_D=\beta=-1$. The symmetry in the functional form of c(\textbf{r}) ensures that the same dynamical steady-states are reached for both chemo-attractive and chemo-repulsive interactions, \textit{i.e.,} hills of the chemical profile would become valleys and vice-versa.
\begin{figure}[t!]
\centering
    \includegraphics[width=1.0\linewidth]{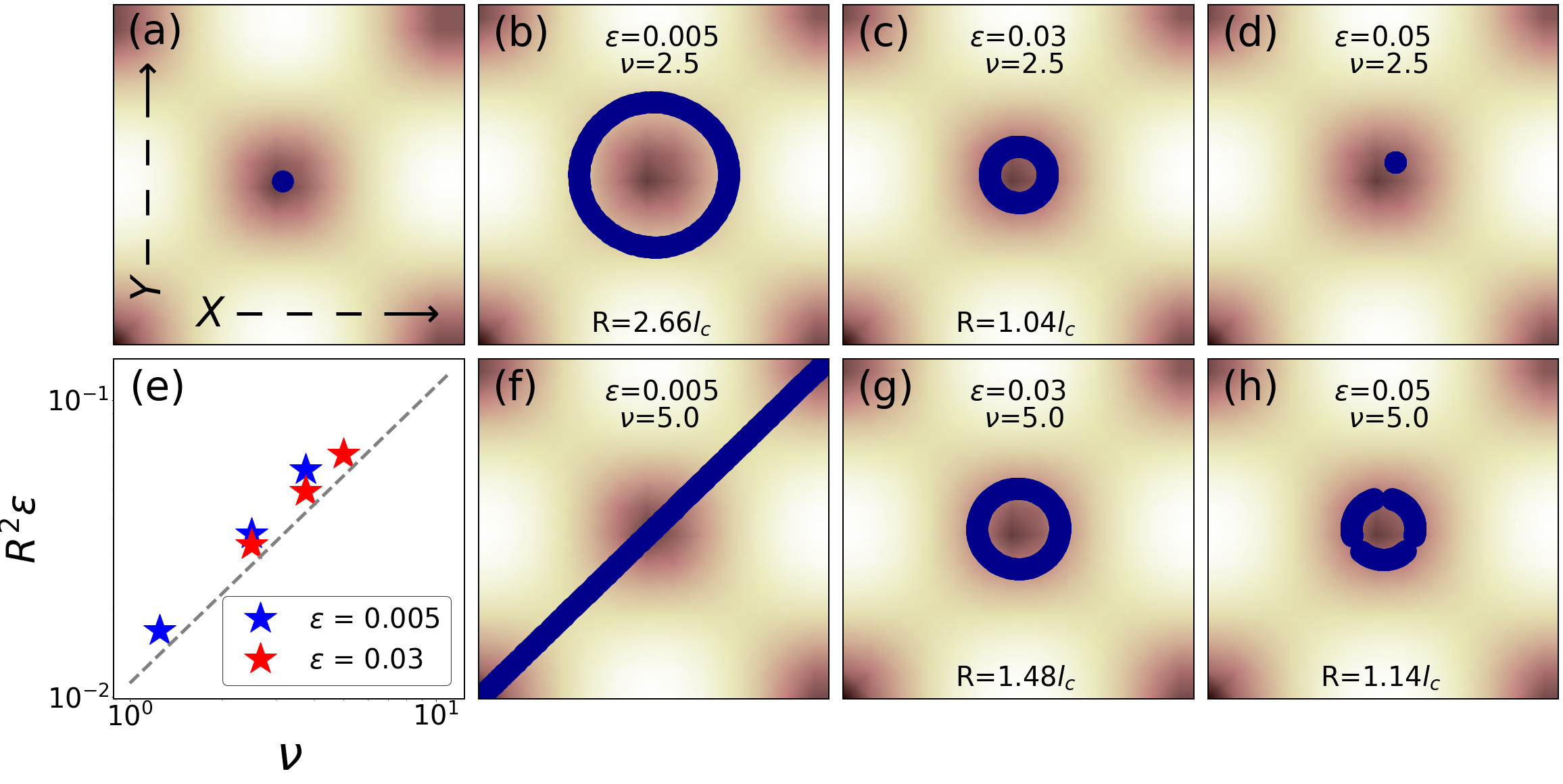}
    \caption{(a) Schematic of the single-particle model. Background color map: \textit{white} for chemical hill, \textit{brown} for chemical valley. Preferred straight-line or circular trajectories of a single particle are shown for times $1500\tau_{c}<t<2500\tau_c$ for some system configurations (b-c) and (f-h). (d) Particle is localized for low $\nu$ and high $\epsilon$. (e) The radius of orbital motion $R$ for circular trajectories increases with $\nu$ and falls with $\epsilon$. The dashed line has a slope of $1$.}
    \label{fig:fig1}
\end{figure}

The scales of the problem are defined using parameters of the substrate: the length up to which a particle translates before it experiences a rotation due to the chemical gradient is given by $l_{c} = \beta_D/\beta$; the ratio $|\beta_D|/\beta^2$ sets an intrinsic time scale $\tau_{c}$. All other times and lengths in the system are scaled with $\tau_c$ and $l_c$. The elastic time scale $(\mu k)^{-1}$ is taken to be $5 \times 10^{-2} \tau_c$. To compare the active force to the chemical force, we define a dimensionless activity $\nu = \frac{v_{0}} {\sqrt{\lambda^{-1} \beta_D \beta}}$ which is varied between $[0.25, 10]$ as $v_0$ $\in [0.05, 2]$. Wavelength is fixed to $\lambda= 25 l_c$. The surface gradient $\epsilon$ = $h_0/\lambda$  quantifies the steepness of chemical concentration and is kept in the range $[10^{-3}, 10^{1}]$. The dynamics and steady state of the system are studied by varying the surface gradient $\epsilon$ which provides the particles an effective attraction to valleys and activity $\nu$ that disperses particles throughout the substrate. Each realization of the system is $5\times$$10^{5} $ time steps long with a time step $\Delta t =5\times$$10^{-3} \tau_{c}$. All statistical quantities are recorded every $10^{3}$ steps. The system is studied for a $L \times L$ square geometry with periodic boundary conditions (PBC) applied in both directions. Data from $20$ independent realizations are used for averaging. {We consider two cases of interest: (A) Single-particle model, (B) Interacting or Many-particle model with a packing density of $\Phi= 0.6$. The chosen density is safely above the critical density $\Phi_{c} = 0.39$ }\cite{prl2012filly} {to produce MIPS but not high enough to cause jamming} \cite{jamming}. The packing density, wavelength, and particle size are crucial in determining the phase diagram. The effect of the variation in these parameters is briefly discussed towards the end of section III.

\section{Results}
\subsection{Single-particle model}
The effect of the patterned surface on the dynamics of a single particle is studied by setting $\mu=0$ and $D_R=0$ in equations \ref{eq3} and \ref{eq4}.  A unit cell of the periodically patterned chemical substrate having size $L = 12.5 l_c$ encompassing exactly one hill of the chemical concentration is chosen. A model cartoon of the cell is shown in Fig. \ref{fig:fig1} (a) and the trajectories are reported in Fig. \ref{fig:fig1} (b-d) for $\nu=2.5$ and (f-h) for $\nu=5$.

For low $\epsilon$ and high $\nu$ [see Fig. \ref{fig:fig1}(f)], the particle exhibits unconfined diffusive dynamics, moving along a straight line. There is a preference to traverse along the $x-y$ direction followed by $x$ and $y$ directions owing to the arrangement of valleys or minima of c(\textbf{r}) and PBC. For a lower $\nu$ or higher $\epsilon$, the particle is unable to escape the chemical valley it was initialized in [see Fig. \ref{fig:fig1}(b-d,g-h)]. The competition between the motility and chemical gradient results in a bias to move on a contour of constant $c$ that is approximately circular with a radius $R$ that decays as $R \sim {(\frac{\nu}{\epsilon})}^{0.5}$ [see Fig. \ref{fig:fig1}(e)]. For a fixed $\nu$, increasing $\epsilon$ leads to steeper chemical profiles and smaller orbital radius $R$ as shown in Fig. \ref{fig:fig1}(b-c,g-h). The confinement becomes stronger for very high $\epsilon$ and low $\nu$ [see Fig. \ref{fig:fig1}(d)] resulting in effective localization of the particle position. 

\begin{figure}[t!]
\centering
    \includegraphics[width=.8\linewidth]{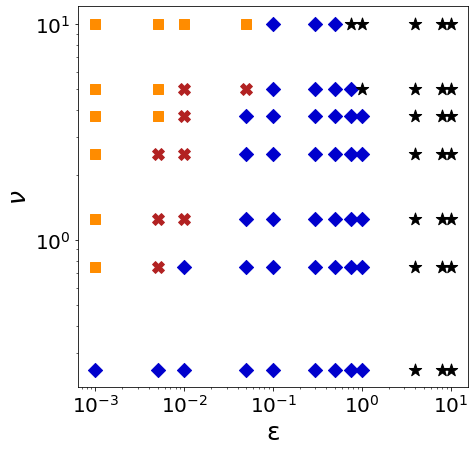} 
    \caption{{State diagram in the $(\nu, \epsilon)$ plane.} Symbols correspond to phases: \textit{square} for \textit{CMIPS}, \textit{cross} for \textit{SRC}, \textit{diamond} for \textit{CC}, \textit{star} for \textit{LC}. 
    }
\label{fig:fig2}
\end{figure}

\subsection{Interacting model}
Particle-particle interaction and noise are turned on by setting $\mu=1$ and $D_{R} = 10^{-4}\tau_{c}^{-1}$ for a system size $L = 100 l_c$. For a packing density $\Phi = \frac{N \times \pi a_0^2}{L \times L}$ of $0.6$, around $\textit{N}\sim 2000$ particles are simulated.  The simulation starts from a homogeneous arrangement of particles with a constant self-propulsion speed and randomized orientations on the substrate. The chemical field dictates the particles to follow a path with lower \textcolor{black}{$c$}, whereas the activity and interaction among the particles encourage them to cluster. Consequently, periodic clusters are formed in the systems for which the surface gradient is non-negligible. The ($\nu$, $\epsilon$) phase-space is explored and the state diagram is presented in Fig. \ref{fig:fig2}. The characteristics of the obtained phases follow. 

\begin{figure}[t!]
\centering
    \includegraphics[width=\linewidth]{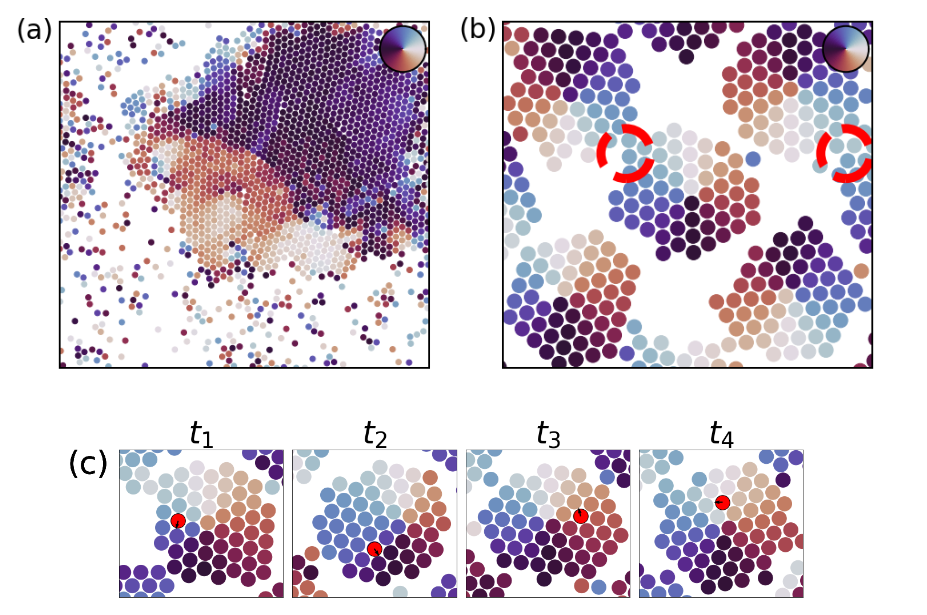}
    \caption{(a) Late-time snapshot of a \textit{CMIPS} system with $\nu=5, \epsilon=0.001$. (b) Zoomed-in late-time snapshot of \textit{SRC} system with $\nu=1.25, \epsilon=0.01$. Red dashed circles indicate areas where particles are exchanged between clusters. (c) Position and orientation of a tagged particle (\textit{red}) revolving in a cluster of \textit{SRC} for times $t_{1}<t_{2}<t_{3}<t_{4}$ separated by $20\tau_{c}$. Particles are colored according to their orientations as indicated by the color ring in the inset.}
    \label{fig:fig3}
\end{figure}

\textit{Chemo-motility-induced phase-separated (CMIPS) state:} For very low $\epsilon \sim 10^{-3}$ and moderate $\nu\geq0.75$, we obtain a rotating macroscopic cluster formation as seen in Fig. \ref{fig:fig3} (a) and Supplementary Material $S1$, wherein a dense liquid phase coexists with the gaseous phase.  The phase separation is due to an interplay between chemo-phoretic interactions that collapse particles into valleys of the chemical concentration forming clusters, and activity that disperses particles from the clusters. This is in contrast with the self-trapping positive feedback that leads to MIPS \cite{prl2012filly, prl2013exp}. \textit{CMIPS} is structurally similar to MIPS which can be retrieved as $\epsilon \to 0$. Unexpectedly, the macroscopic clustering found for \textit{CMIPS} phase appears at much lower dimensionless activity $\nu \sim 0.75$, whereas for the same packing density of particles on a flat surface, the minimum P\'eclet number $Pe = \frac{v_0}{D_R a_0}$ required to obtain MIPS is much larger $\sim 50-100$ \cite{peclet}. Some key differences between \textit{CMIPS} and MIPS are shown in Fig. \ref{fig:fig6}(a) and \ref{fig:fig5}(a). Unlike in MIPS, the presence of small local gradients \textit{CMIPS} weakens the macroscopic clustering and orientational correlations and slows down the translational dynamics of particles resulting in lowered diffusivity.

\textit{Synchronized Rotating clusters (SRC)}: For slightly higher $\epsilon \sim 10^{-2}$ and moderate $\nu$, the periodic nature of the chemical concentration is seen through the formation of periodic clusters.  The macroscopic \textit{CMIPS} phase is suppressed as strong chemical gradients induce local splitting. \textcolor{black}{These periodic clusters are composed of orbits of particles that revolve around the cluster center} as shown in Fig. \ref{fig:fig3} (b). A sense of rotation is picked up by a cluster without any alignment mechanism such that any two nearest clusters at a point in time rotate oppositely. The handedness of a cluster may also change with time, and this change is reciprocated by the rest of the system [refer Supplementary Material $S2$]. This is possible because these clusters are spatially well-connected and can exchange particles. 

\begin{figure}[t!]
\centering
    \includegraphics[width=1.0\linewidth]{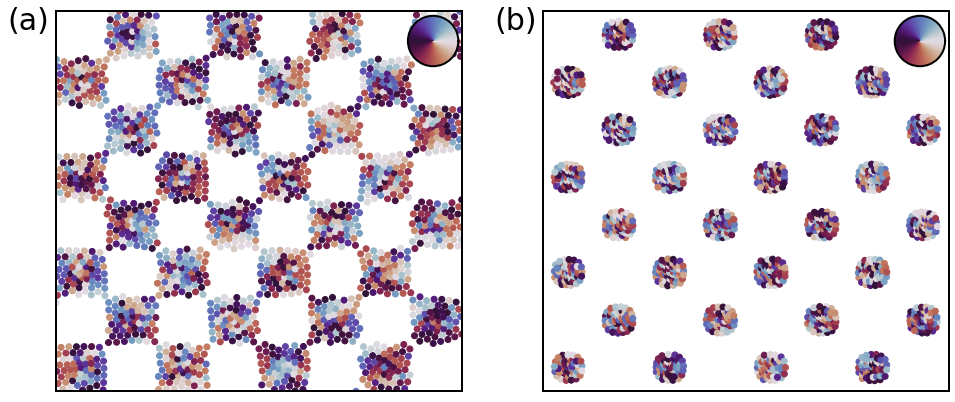} 
    \caption{Late-time snapshots for (a) \textit{CC} system with $\nu=3.75, \epsilon=1$ and (b) \textit{LC} system with $\nu=3.75, \epsilon=10$. Particles are colored according to their orientations as indicated by the color ring in the inset.}
    \label{fig:fig3c}
\end{figure}

\textit{Localized clusters (\textit{LC}):} For very high $\epsilon\geq10^{0}$, particle trajectories asymptotically converge to bounded areas in space leading to trapping in the valleys of the chemical profile [see Fig. \ref{fig:fig3c} (b-c)]. The dynamics of one cluster are independent of the others in the system. Connectivity between the clusters is lost as particles overlap with each other, and the system collapses to the non-interacting model, \textit{i.e.,} each localized cluster can be treated as an independent subsystem. {This phase arises purely due to the employment of soft excluded volume interactions in the model. Were the interactions hard, then in this range of $\epsilon$, particles would randomly hop from one localized region of the system to another.} 

\textit{Crossover clusters (CC):} For $\epsilon$ ranging between the \textit{SRC} and \textit{LC} phases, \textit{i.e.,} $\epsilon \sim 10^{-1}$, periodic clusters are formed  as shown in Fig. \ref{fig:fig3c} (a). These clusters serve as the cross-over phase between \textit{SRC} and \textit{LC} phases. They are characterized by \textit{hopping transport} of particles from one cluster to another but occasionally may show non-synchronized incomplete rotations of a few clusters of the system, depending on the system parameters. Each cluster acts like a chemo-repulsive shell due to the local anisotropy in the chemical concentration. This constricts the freedom of a cluster to grow beyond a certain size. The details of such dynamics are discussed later.

\begin{figure}[t!]
\centering
    \includegraphics[width=0.8\linewidth]{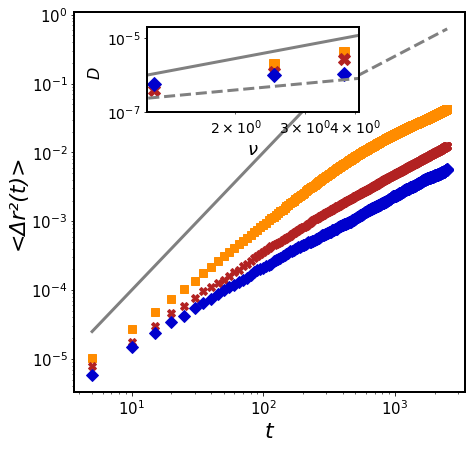} 
    \caption{Mean-square displacement $<\Delta r^2 (t)>$ \textit{vs.} time for \textit{CMIPS, SRC}, and \textit{CC} phases showing crossover from ballistic to diffusive regime. Inset shows diffusivity \textit{D} as a function of \textit{$\nu$}. The dashed line is drawn for slope $1$ and the solid line is drawn for slope $2$. Key: \textit{orange square} for \textit{CMIPS} $(\nu, \epsilon) = (3.75,0.001)$, \textit{red cross} for \textit{SRC} $(2.5, 0.01)$ and \textit{blue diamond} for \textit{CC} $(3.75,0.1)$
    }
    \label{fig:fig4}
\end{figure}

\begin{figure}[t!]
\centering
    \includegraphics[width=0.8\linewidth]{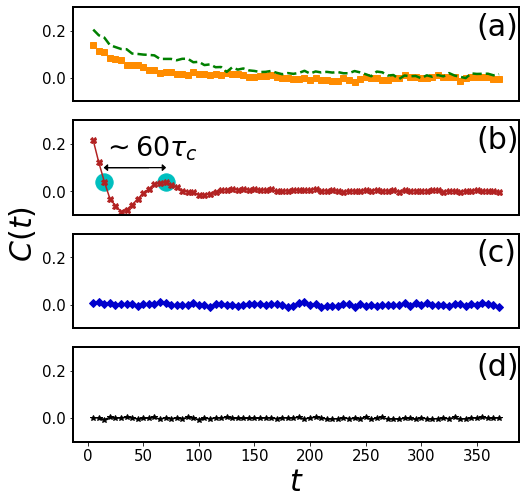} 
    \caption{Variation of the particle orientational auto-correlation function $C(t)$ for (a) {\textit{CMIPS} and MIPS (\textit{green})} (b) \textit{SRC} (c) \textit{CC} and (d) \textit{LC} phases are plotted for $350\tau_{c}$ in the steady state. Long-tailed decay in (a) and oscillations in (b) indicate the presence of collective rotations. The system configurations for the four phases are the same as in Fig. \ref{fig:fig4}.}
    \label{fig:fig6}
    
\end{figure}

\begin{figure}[t!]
\centering
    \includegraphics[width=0.8\linewidth]{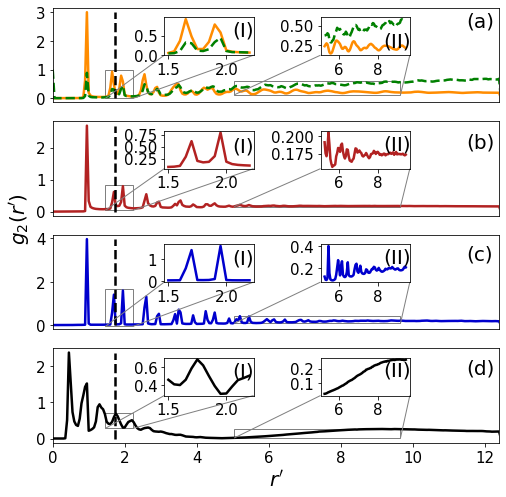} 
    \caption{The pair correlation function g$_{2}(r')$ for (a) {\textit{CMIPS} and MIPS (\textit{green})} (b) \textit{SRC} (c) \textit{CC} and (d) \textit{LC} phases. The \textit{black} dashed line is drawn at $r'=\sqrt{3}$. Positions of peaks throw light on structural ordering and overlap between particles in the respective phases. Inset I zooms into $r'\in(1.5,2.25)$. Inset II zooms into $r'\in(5.25,9.5)$. The system parameters for the four phases are the same as in Fig. \ref{fig:fig4}. }
    \label{fig:fig5}
\end{figure}

The dynamics of the phases are quantified by calculating the mean square displacement (MSD) of the particles: 
\begin{equation}
\end{equation}
where $<..>$ stands for average over many reference times $t_0$, particles and realisations. MSD regime shifts from ballistic (slope $2$) during initial times to diffusive (slope $1$) for late times as shown in Fig. \ref{fig:fig4}. The MSD of particles systematically decreases as we go from \textit{CMIPS} to \textit{CC} phase. The effective diffusivity $\textit{D}=\lim_{t \to \infty}<\Delta r^2(t)>/4t$ is shown in the inset of Fig. \ref{fig:fig4}. {$D \in [10^{-8}, 10^{-5}]$ for \textit{CMIPS}; $D \in [10^{-9}, 10^{-8}]$ for \textit{SRC}; $D \in [10^{-11}, 10^{-8}]$ for \textit{CC}.} We find the scaling relation: \textit{D} $\sim$ $\nu$$^{\beta}$. We obtain $\beta \simeq 2$ for {\it CMIPS} as is known for other active systems \cite{beta1, beta2}; $1<\beta<2$ for \textit{SRC}; $0<\beta<2$ for \textit{CC}. \textit{D} is zero for the \textit{LC} phase irrespective of $\nu$. The repressed translational dynamics (note the low diffusivities) in \textit{CMIPS} and \textit{SRC} phases are compensated by enhanced rotational dynamics, as discussed next.

Orientational dynamics of particles are characterized by measuring the {orientational auto-correlation function (OACF)}: 
\begin{equation}
C(t) = <\cos(\phi_{i}(t_{0})-\phi_{i}(t+t_{0}))>-<\cos\phi_{i}(t+t_{0})>^{2}
\end{equation}
where $<...> $ has the same meaning as before. {$\phi_{i}$ is the direction of the instantaneous velocity of the $i^{th}$ particle.} OACF exponentially decays for \textit{CMIPS} [see Fig. \ref{fig:fig6} (a)] with a decay time of $t_{c}\sim 50\tau_{c}$. OACF of MIPS for the same self-propulsion speed is found to have a correlation time twice as large. This renders the conclusion that a finite $\epsilon$ induces local splitting of clusters disturbing the orientation correlations. Orientations of particles in the \textit{CC} and \textit{LC} phases do not share a functional relationship, leading to a zero OACF [see Fig. \ref{fig:fig6} (c-d)].

The OACF shows clear oscillations for the \textit{SRC} phase indicating robust rotational order in the system [see Fig. \ref{fig:fig6} (b)]. To support it, snapshots of a single cluster with a \textit{red} tagged particle and its variation over time are shown in Fig. \ref{fig:fig3} (c). The time taken to complete one full cycle is annotated in Fig. \ref{fig:fig6} (b) by highlighting two orientations separated in time by $t_{4}-t_{1}=60\tau_{c}$. While it is the steepness of the valley ($\epsilon$) that drives the particles into periodic clusters, once the clusters are formed, the activity alone ensures the particle dynamics inside the cluster. However, moving tangentially on orbits of fixed radii from the center of the unit cell encompassing a valley is the only means to minimize the surface potential. This leads to an almost circular trajectory of the particles. Note that for a similar range of $\epsilon$, the trajectory of a single particle is also circular with finite radius as shown in Fig. \ref{fig:fig1}. 

The handedness of any two nearest rotating clusters in \textit{SRC} is opposite [see Fig. \ref{fig:fig3} (b)]. The sense of rotation of a cluster is purely decided by the particles that are at the outermost orbit as they have the highest magnitude of instantaneous velocity in the cluster. The regions where particles are exchanged between clusters are highlighted in \textit{red} dashed circles. A particle from a cluster leaving to join one of the nearest clusters retains its sense of motion as it is energetically unfeasible to reorient it opposite to its original propelling direction. This gives rise to a specific sense of rotational order in all the clusters such that any two nearest clusters rotate oppositely. In this way, activity, steric repulsion, and periodicity of the chemical profile lead to synchronized rotations in the whole system. 

The structural ordering of the particles in different phases is characterized by the radial distribution function (RDF) $g_{2}$(r). RDF is a measure of the probability of finding a particle at $\textbf{r}_{2}$ given a particle at $\textbf{r}_{1}$ with $r=|\textbf{r}_{2}$-$\textbf{r}_{1}|$. In two dimensions, $<n>g(r)d^{2}r$ gives the number of particles in $d^{2}r$, where $<n>$ is the mean number of particles in the unit area. RDF is plotted against the normalized radial distance $r' = r/(2a_{0})$ in Fig. \ref{fig:fig5}. Evidently, \textit{CMIPS}, \textit{SRC}, and \textit{CC} show their largest peak at the nearest-neighbor (\textit{nn}) distance $r'=1$. The second and third peaks occur at  $r'=\sqrt{3}$ (second \textit{nn}) and $r'=2$ (third \textit{nn}) respectively [see insets I of Fig. \ref{fig:fig5}]. This indicates the presence of hexagonal close-packing (HCP). {We also compared the RDF of \textit{CMIPS} with a MIPS system having the same $v_0$ in Fig.} \ref{fig:fig5}{(a). We find that the location of the first few peaks is the same for both cases, but the peaks are sharper in \textit{CMIPS} indicating tighter packing than MIPS. Also, at larger radial distances, RDF of \textit{CMIPS} approaches a value lower than $1$ due to the voids created by the maxima of the chemical profile, while RDF of MIPS approaches $1$ indicating a random arrangement of particles.} 

For large $\epsilon > 10^0$, the particles descend towards the minima of {$c$} after landing on a cell and are unable to escape the minima. As the chemical force exceeds the steric interaction and active force, the soft nature of the steric force allows particles to overlap, shifting the peaks of $g_2(r)$ towards lower radial distances [see Fig. \ref{fig:fig5} (d)]. This localizes particle positions resulting in zero diffusivity. Insets II of Fig. \ref{fig:fig5} zoom into the radial distances near the start of the next periodic valley. \textit{CMIPS} shows long-range ordering, \textit{LC} indicates periodicity in clustering but such information is inconclusive in the case of \textit{SRC} and \textit{CC}.

\begin{figure}[tb!]
\centering
    \includegraphics[width=1.0\linewidth]{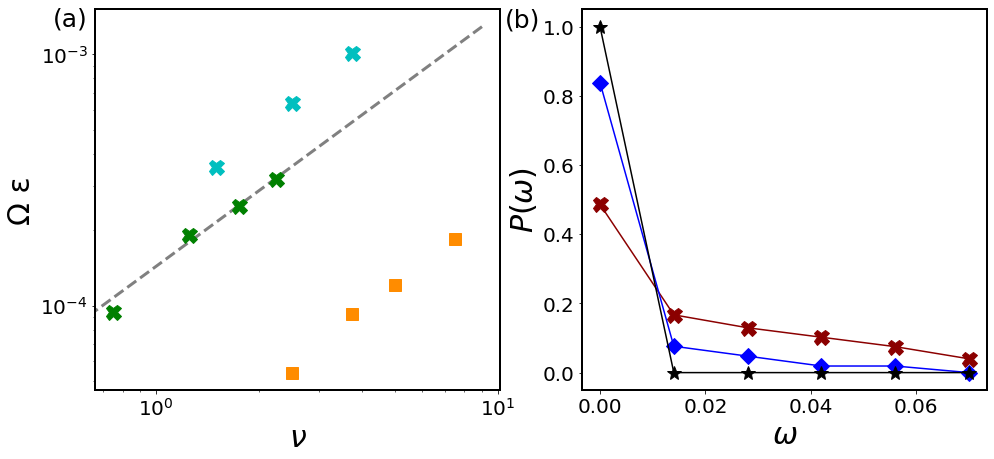} 
    \caption{{(a) The global angular frequency $\Omega(\nu,\epsilon)$ increases linearly with $\nu$ for \textit{SRC} and \textit{CMIPS} systems. \textit{Key}: \textit{cyan}- $\epsilon=0.01$, \textit{green}- $\epsilon=0.005$, and \textit{orange}- $\epsilon=0.001$. The dashed line has a slope of $1$. (b) The local angular frequency $\omega(\nu,\epsilon)$ for a chosen \textit{SRC} $(\nu, \epsilon) = (0.25, 0.01)$, \textit{CC} ($0.5, 0.5$), and \textit{LC} ($0.5, 10$) system. Symbols have the same meaning as in Fig.} \ref{fig:fig2}.}
    \label{fig:fig9}
\end{figure}

The macroscopic cluster in \textit{CMIPS} also rotates as a part or whole [refer Supplementary Material $S1$]. This may occur as a result of the cluster boundary being atop chemical maxima that provides the particles with a reorienting torque and force to move away from the maxima. Hence, our rotational states come from both \textit{CMIPS} (macroscopic rotations) and \textit{SRC} (synchronized rotations) phases. To characterize the strength of collective rotations we estimate the global angular frequency $\Omega(\nu,\epsilon)$: 
\begin{equation}
    \Omega (\nu, \epsilon) = <\frac{1}{N_{c}}\sum_{i=1}^{N_{c}} |\sum_{j=1}^{N_{c, j}} \frac{(\textbf{r}_{j}\times \textbf{v}_{j})}{r_{j}^{2}}| >
\end{equation}
 where $N_{c}$ denotes the number of valleys in the system (fixed for a certain $\lambda$); $N_{c, j}$ refers to the number of particles in the $i^{th}$ cluster computed by counting particles within a radial distance of $\lambda/8$ from the center of the $i^{th}$ valley. $j^{th}$ particle has position ${\bf r}_j$ relative to the center of the valley and instantaneous velocity ${\bf v}_j$. A scaled $\Omega = \epsilon \Omega$ is chosen to filter the effect of activity on global rotation. Scaled $\Omega$ increases linearly with $\nu$ proving that clusters with high activity rotate faster as shown in Fig. \ref{fig:fig9}(a). $\Omega$ is also comparable for \textit{SRC} and \textit{CMIPS} for the same activity.
 
To understand the variation from \textit{SRC} to \textit{LC}, we measure the probability distribution function of local angular frequency $P(\omega)$, where $\omega = <|\sum_{j=1}^{N_{c, j}} \frac{(\textbf{r}_{j}\times \textbf{v}_{j})}{r_{j}^{2}}|>$. This distribution is obtained over all the unit cells, many time snapshots and realizations in the steady state. In Fig. \ref{fig:fig9}(b) we show the plot of $P(\omega)$ vs. $\omega$ for \textit{SRC}, \textit{CC}, and \textit{LC} phase. \textit{SRC} phase shows a broad distribution with a large tail, a signature of robust rotations. The \textit{LC} phase shows a delta function with a peak at $\omega=0$ as it has no rotations. The \textit{CC} phase shows a short tail with finite moderate frequencies proving its dual behavior with both localized particles and incomplete rotations depending on the parameter space $(\nu, \epsilon)$.
    
The state diagram and location of the phases also depend on packing density. For example, for a smaller packing density of $\Phi=0.45$ with $\nu=2.5$ and $\lambda=25l_c$, \textit{CMIPS} is only observed very close to $\epsilon=0$ and other phases are ill-defined. Since the synchronized rotations in the \textit{SRC} phase can be obtained only if all the local clusters are well-connected, a packing density larger than the critical packing density of disks in two-dimensions $\Phi > 0.6$ \cite{percolation} is required. The relation between particle size and the wavelength $\lambda$ is also crucial. For wavelength $\lambda$ larger than particle size $a_0$, many particles participate in forming a cluster and hence an HCP structure is achieved. If both parameters lie in the same range, we may not be able to observe the exchange of particles in the $SRC$ phase or achieve a dense packing of HCP. On the other hand, if the particle size is large compared to $\lambda$, the particles do not sense the chemical patterns well and move like they would on a flat surface. 

\begin{figure}[t!]
\centering
    \includegraphics[width=0.8\linewidth]{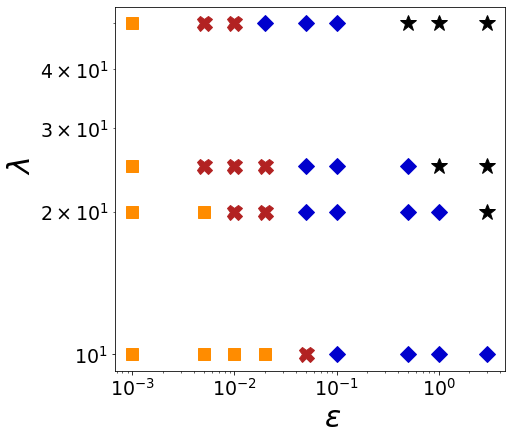} 
    \caption{{Phase diagram in the $(\lambda,\epsilon)$ phase space is shown for packing density $\Phi=0.6$ and $\nu=2.5$. The chosen values of $\lambda$ correspond to $10 l_c$, $20 l_c$, $25 l_c$, and $50 l_c$. Small $\lambda$ promotes the formation of \textit{CMIPS} and extinguishes \textit{SRC}. Large $\lambda$ suppresses both \textit{CMIPS} and \textit{SRC}. Symbols have the same meaning as in Fig.} \ref{fig:fig2}.}
    \label{fig:fig10}
\end{figure}

We further show the effect of varying the wavelength $\lambda$ of the chemical profile in Fig. \ref{fig:fig10} keeping other parameters fixed. For high $\lambda$, consecutive chemical valleys are spread further apart in space, making the crossing of peaks difficult and conditions unfavorable to any macroscopic cluster formation and collective rotation. Such a profile suppresses \textit{CMIPS} and \textit{SRC}, making the phase boundaries shift to the left. For low $\lambda$, the large periodicity of the chemical profile increases the uniformity of the substrate and makes it favorable for \textit{CMIPS} to exist for a larger range of parameters but the \textit{SRC} is shrunk due to fewer number of particles in each local cell. This shifts the phase boundary of \textit{CMIPS} to the right and extinguishes the \textit{SRC} phase. Again for larger $\epsilon$, the system always goes to the \textit{LC} phase due to soft-exclusion amongst the particles. 

\section{Discussion}

Dynamics and steady states of a collection of chemo-phoretically interacting ABPs are studied for the case of a chemical potential quenched in time and periodic in space. The work elucidates competing interests between activity and chemophoresis. In the extreme limits, when activity dominates we obtain chemo-MIPS \textit{i.e., CMIPS} - a macroscopic rotating cluster, and when gradients dominate we obtain localized clusters with glassy dynamics. However, when the two forces are comparable, particles arrange themselves into periodic clusters of finite length showing synchronized rotations spanning the whole system. The macroscopic clusters obtained here for the \textit{CMIPS} phase are very similar to MIPS obtained for much larger activity with some characteristic differences of strong collective rotation, which appears at the cost of suppressed translational diffusion. The presence of local gradients eases the self-trapping mechanism of \textit{CMIPS} and encourages cluster formation even at smaller activities.

The mechanism of the formation of the macroscopic cluster in \textit{CMIPS} is distinct from the liquid-vapor phase separation observed in equilibrium for attractive particles \cite{equilibiurmliquidvapour} where the long-range attractive interaction among the particles is responsible for the formation of liquid-vapor phase separation in equilibrium. However, in \textit{CMIPS} an effective attraction towards the local minima of the chemical profile motivates the formation of macroscopic clusters. The clusters in \textit{CMIPS} are very dynamic and have characteristic global rotation, whereas the clusters of the liquid phase in equilibrium are dynamic only due to thermal diffusion. 

Fascinatingly, in the case of synchronized rotations, a strict sense of handedness is picked up by a cluster without any intrinsic alignment interaction within the model. An interplay of time-reversal asymmetry and chemo-phoretic interactions between the repulsive disks is responsible for such collective rotations in the system. These collective responses can be easily reproduced for various types of taxis, viz. phototaxis \cite{photo}, viscotaxis \cite{visco}, electrotaxis \cite{electro}, thermotaxis \cite{thermo}, etc. The observed rotations are more robust and are in contrast with swarms that generally have one cluster rotating about its center of mass \cite{swarm} as a response to external obstacles or phoretic-motility \cite{honeybee, swarming}. These rotations can even be exploited by designing mechanical devices to generate torques using active particles.

The chemically-induced rotational tendency we observe in this study is very different from the rotating clusters that appear when particles are characterized by an attractive interaction and are elongated \cite{paper2}. In such cases, a competition between the short-range attraction, long-range repulsion, and the asymmetric shape of particles introduces an extra torque effect in the particle dynamics and leads to the rotating phase. However, in our study, rotation is observed for symmetric particles solely due to competition between steric repulsion and the activity of the particles. They are very similar to the global rotations found in a recent study \cite{vicseknatphys}. 

Phase separation can give rise to interesting dynamics that can be the result of the instabilities present in the coarse-grained equations of density and local polarization of particles \cite{tonertupre1998, catesprl2013}. Such studies may provide a description of dynamical heterogeneity and fluctuations. A similar approach is adopted in \cite{paper1} to show clustering and pattern formation in chemorepulsive active colloids. Other recent simulation techniques to characterize similar investigations include the kinetic Monte Carlo method \cite{KMC} and Molecular dynamics \cite{NEMD}. Our study gives a microscopic understanding of the rotation of a collection of active particles in many natural systems and can be tested by designing experimental setups involving microswimmers on patterned substrates. Such experiments can be crucial in understanding the collective response of biological swimmers to the underlying medium. In future work, it would be interesting to study a model that embodies opposite signs of the chemical coupling coefficients or spatially random or time-dependent taxis.

\section{Conflicts of interest} 
There are no conflicts of interest to declare.

\section{Acknowledgements} 
P.E. acknowledges the support and the resources provided by PARAM Shivay Facility under the National Supercomputing Mission, Government of India at the Indian Institute of Technology, Varanasi. S.M. thanks DST-SERB India, MTR/2021/000438, and CRG/2021/006945 for financial support.

\section{Author Contributions} 
The problem was designed by S.M. and numerically investigated by P.E. Both authors analyzed and interpreted the results. The manuscript was prepared by P.E. Both authors approved the final version of the manuscript. 

\bibliographystyle{apsrev4-1}
\bibliography{reference}

\begin{thebibliography}{58}%
\makeatletter
\providecommand \@ifxundefined [1]{%
 \@ifx{#1\undefined}
}%
\providecommand \@ifnum [1]{%
 \ifnum #1\expandafter \@firstoftwo
 \else \expandafter \@secondoftwo
 \fi
}%
\providecommand \@ifx [1]{%
 \ifx #1\expandafter \@firstoftwo
 \else \expandafter \@secondoftwo
 \fi
}%
\providecommand \natexlab [1]{#1}%
\providecommand \enquote  [1]{``#1''}%
\providecommand \bibnamefont  [1]{#1}%
\providecommand \bibfnamefont [1]{#1}%
\providecommand \citenamefont [1]{#1}%
\providecommand \href@noop [0]{\@secondoftwo}%
\providecommand \href [0]{\begingroup \@sanitize@url \@href}%
\providecommand \@href[1]{\@@startlink{#1}\@@href}%
\providecommand \@@href[1]{\endgroup#1\@@endlink}%
\providecommand \@sanitize@url [0]{\catcode `\\12\catcode `\$12\catcode `\&12\catcode `\#12\catcode `\^12\catcode `\_12\catcode `\%12\relax}%
\providecommand \@@startlink[1]{}%
\providecommand \@@endlink[0]{}%
\providecommand \url  [0]{\begingroup\@sanitize@url \@url }%
\providecommand \@url [1]{\endgroup\@href {#1}{\urlprefix }}%
\providecommand \urlprefix  [0]{URL }%
\providecommand \Eprint [0]{\href }%
\providecommand \doibase [0]{http://dx.doi.org/}%
\providecommand \selectlanguage [0]{\@gobble}%
\providecommand \bibinfo  [0]{\@secondoftwo}%
\providecommand \bibfield  [0]{\@secondoftwo}%
\providecommand \translation [1]{[#1]}%
\providecommand \BibitemOpen [0]{}%
\providecommand \bibitemStop [0]{}%
\providecommand \bibitemNoStop [0]{.\EOS\space}%
\providecommand \EOS [0]{\spacefactor3000\relax}%
\providecommand \BibitemShut  [1]{\csname bibitem#1\endcsname}%
\let\auto@bib@innerbib\@empty
\bibitem [{\citenamefont {Bechinger}\ \emph {et~al.}(2016)\citenamefont {Bechinger}, \citenamefont {Leonardo}, \citenamefont {Löwen}, \citenamefont {Reichhardt}, \citenamefont {Volpe},\ and\ \citenamefont {Volpe}}]{rmpbechinger2016}%
  \BibitemOpen
  \bibfield  {author} {\bibinfo {author} {\bibfnamefont {C.}~\bibnamefont {Bechinger}}, \bibinfo {author} {\bibfnamefont {R.~D.}\ \bibnamefont {Leonardo}}, \bibinfo {author} {\bibfnamefont {H.}~\bibnamefont {Löwen}}, \bibinfo {author} {\bibfnamefont {C.}~\bibnamefont {Reichhardt}}, \bibinfo {author} {\bibfnamefont {G.}~\bibnamefont {Volpe}}, \ and\ \bibinfo {author} {\bibfnamefont {G.}~\bibnamefont {Volpe}},\ }\href@noop {} {\bibfield  {journal} {\bibinfo  {journal} {Rev. Mod. Phys. \textbf{88}, 045006}\ } (\bibinfo {year} {2016})}\BibitemShut {NoStop}%
\bibitem [{\citenamefont {Toner}\ \emph {et~al.}(2005)\citenamefont {Toner}, \citenamefont {Tu},\ and\ \citenamefont {Ramaswamy}}]{annalsofphys2005}%
  \BibitemOpen
  \bibfield  {author} {\bibinfo {author} {\bibfnamefont {J.}~\bibnamefont {Toner}}, \bibinfo {author} {\bibfnamefont {Y.}~\bibnamefont {Tu}}, \ and\ \bibinfo {author} {\bibfnamefont {S.}~\bibnamefont {Ramaswamy}},\ }\href@noop {} {\bibfield  {journal} {\bibinfo  {journal} {Ann. Phys. (N.Y.), \textbf{318}, 170}\ } (\bibinfo {year} {2005})}\BibitemShut {NoStop}%
\bibitem [{\citenamefont {Vicsek}\ and\ \citenamefont {Zafeiris}(2012)}]{vicsek2012}%
  \BibitemOpen
  \bibfield  {author} {\bibinfo {author} {\bibfnamefont {T.}~\bibnamefont {Vicsek}}\ and\ \bibinfo {author} {\bibfnamefont {A.}~\bibnamefont {Zafeiris}},\ }\href@noop {} {\bibfield  {journal} {\bibinfo  {journal} {Phys. Rep. \textbf{517}, 71}\ } (\bibinfo {year} {2012})}\BibitemShut {NoStop}%
\bibitem [{\citenamefont {Marchetti}\ \emph {et~al.}(2013)\citenamefont {Marchetti}, \citenamefont {Joanny}, \citenamefont {Ramaswamy}, \citenamefont {Liverpool}, \citenamefont {Prost}, \citenamefont {Rao},\ and\ \citenamefont {Simha}}]{rmp2013}%
  \BibitemOpen
  \bibfield  {author} {\bibinfo {author} {\bibfnamefont {M.~C.}\ \bibnamefont {Marchetti}}, \bibinfo {author} {\bibfnamefont {J.~F.}\ \bibnamefont {Joanny}}, \bibinfo {author} {\bibfnamefont {S.}~\bibnamefont {Ramaswamy}}, \bibinfo {author} {\bibfnamefont {T.~B.}\ \bibnamefont {Liverpool}}, \bibinfo {author} {\bibfnamefont {J.}~\bibnamefont {Prost}}, \bibinfo {author} {\bibfnamefont {M.}~\bibnamefont {Rao}}, \ and\ \bibinfo {author} {\bibfnamefont {R.~S.}\ \bibnamefont {Simha}},\ }\href@noop {} {\bibfield  {journal} {\bibinfo  {journal} {Rev. Mod. Phys. \textbf{85}, 1143}\ } (\bibinfo {year} {2013})}\BibitemShut {NoStop}%
\bibitem [{\citenamefont {Poujade}\ \emph {et~al.}(2007)\citenamefont {Poujade}, \citenamefont {Grasland-Mongrain}, \citenamefont {Hertzog}, \citenamefont {J.~Jouanneau}, \citenamefont {Ladoux}, \citenamefont {Buguin},\ and\ \citenamefont {Silberzan}}]{cell}%
  \BibitemOpen
  \bibfield  {author} {\bibinfo {author} {\bibfnamefont {M.}~\bibnamefont {Poujade}}, \bibinfo {author} {\bibfnamefont {E.}~\bibnamefont {Grasland-Mongrain}}, \bibinfo {author} {\bibfnamefont {A.}~\bibnamefont {Hertzog}}, \bibinfo {author} {\bibfnamefont {P.~C.}\ \bibnamefont {J.~Jouanneau}}, \bibinfo {author} {\bibfnamefont {B.}~\bibnamefont {Ladoux}}, \bibinfo {author} {\bibfnamefont {A.}~\bibnamefont {Buguin}}, \ and\ \bibinfo {author} {\bibfnamefont {P.}~\bibnamefont {Silberzan}},\ }\href@noop {} {\bibfield  {journal} {\bibinfo  {journal} {PNAS \textbf{104}, 15988–15993}\ } (\bibinfo {year} {2007})}\BibitemShut {NoStop}%
\bibitem [{\citenamefont {Feder}(2007)}]{physicstoday2006}%
  \BibitemOpen
  \bibfield  {author} {\bibinfo {author} {\bibfnamefont {T.}~\bibnamefont {Feder}},\ }\href@noop {} {\bibfield  {journal} {\bibinfo  {journal} {Physics Today \textbf{60}, 28}\ } (\bibinfo {year} {2007})}\BibitemShut {NoStop}%
\bibitem [{\citenamefont {Helbing}\ \emph {et~al.}(2000)\citenamefont {Helbing}, \citenamefont {Farkas},\ and\ \citenamefont {Vicsek}}]{humancrowd1}%
  \BibitemOpen
  \bibfield  {author} {\bibinfo {author} {\bibfnamefont {D.}~\bibnamefont {Helbing}}, \bibinfo {author} {\bibfnamefont {I.}~\bibnamefont {Farkas}}, \ and\ \bibinfo {author} {\bibfnamefont {T.}~\bibnamefont {Vicsek}},\ }\href@noop {} {\bibfield  {journal} {\bibinfo  {journal} {Nature \textbf{407}, 487–490}\ } (\bibinfo {year} {2000})}\BibitemShut {NoStop}%
\bibitem [{\citenamefont {Bain}\ and\ \citenamefont {Bartolo}(2019)}]{humancrowd2}%
  \BibitemOpen
  \bibfield  {author} {\bibinfo {author} {\bibfnamefont {N.}~\bibnamefont {Bain}}\ and\ \bibinfo {author} {\bibfnamefont {D.}~\bibnamefont {Bartolo}},\ }\href@noop {} {\bibfield  {journal} {\bibinfo  {journal} {Science \textbf{363}, 46–49}\ } (\bibinfo {year} {2019})}\BibitemShut {NoStop}%
\bibitem [{\citenamefont {Schloesser}\ \emph {et~al.}(2021)\citenamefont {Schloesser}, \citenamefont {Hollenbeck},\ and\ \citenamefont {Kello}}]{foraging}%
  \BibitemOpen
  \bibfield  {author} {\bibinfo {author} {\bibfnamefont {D.~S.}\ \bibnamefont {Schloesser}}, \bibinfo {author} {\bibfnamefont {D.}~\bibnamefont {Hollenbeck}}, \ and\ \bibinfo {author} {\bibfnamefont {C.~T.}\ \bibnamefont {Kello}},\ }\href@noop {} {\bibfield  {journal} {\bibinfo  {journal} {Sci Rep \textbf{11}, 8492}\ } (\bibinfo {year} {2021})}\BibitemShut {NoStop}%
\bibitem [{\citenamefont {Vicsek}\ \emph {et~al.}(1995)\citenamefont {Vicsek}, \citenamefont {Czirók}, \citenamefont {Ben-Jacob}, \citenamefont {Cohen},\ and\ \citenamefont {Shochet}}]{bacteria1}%
  \BibitemOpen
  \bibfield  {author} {\bibinfo {author} {\bibfnamefont {T.}~\bibnamefont {Vicsek}}, \bibinfo {author} {\bibfnamefont {A.}~\bibnamefont {Czirók}}, \bibinfo {author} {\bibfnamefont {E.}~\bibnamefont {Ben-Jacob}}, \bibinfo {author} {\bibfnamefont {I.}~\bibnamefont {Cohen}}, \ and\ \bibinfo {author} {\bibfnamefont {O.}~\bibnamefont {Shochet}},\ }\href@noop {} {\bibfield  {journal} {\bibinfo  {journal} {Phys. Rev. Lett., \textbf{75}, 1226–1229}\ } (\bibinfo {year} {1995})}\BibitemShut {NoStop}%
\bibitem [{\citenamefont {Chaté}(2020)}]{bacteria2}%
  \BibitemOpen
  \bibfield  {author} {\bibinfo {author} {\bibfnamefont {H.}~\bibnamefont {Chaté}},\ }\href@noop {} {\bibfield  {journal} {\bibinfo  {journal} {Annu. Rev. Condens. Mat., \textbf{11}, 189–192}\ } (\bibinfo {year} {2020})}\BibitemShut {NoStop}%
\bibitem [{\citenamefont {Liebchen}\ and\ \citenamefont {Levis}(2017)}]{bacteria3}%
  \BibitemOpen
  \bibfield  {author} {\bibinfo {author} {\bibfnamefont {B.}~\bibnamefont {Liebchen}}\ and\ \bibinfo {author} {\bibfnamefont {D.}~\bibnamefont {Levis}},\ }\href@noop {} {\bibfield  {journal} {\bibinfo  {journal} {Phys. Rev. Lett., \textbf{119}, 058002}\ } (\bibinfo {year} {2017})}\BibitemShut {NoStop}%
\bibitem [{\citenamefont {Theurkauff}\ \emph {et~al.}(2012)\citenamefont {Theurkauff}, \citenamefont {Cottin-Bizonne}, \citenamefont {Palacci}, \citenamefont {Ybert},\ and\ \citenamefont {Bocquet}}]{activecolloids}%
  \BibitemOpen
  \bibfield  {author} {\bibinfo {author} {\bibfnamefont {I.}~\bibnamefont {Theurkauff}}, \bibinfo {author} {\bibfnamefont {C.}~\bibnamefont {Cottin-Bizonne}}, \bibinfo {author} {\bibfnamefont {J.}~\bibnamefont {Palacci}}, \bibinfo {author} {\bibfnamefont {C.}~\bibnamefont {Ybert}}, \ and\ \bibinfo {author} {\bibfnamefont {L.}~\bibnamefont {Bocquet}},\ }\href@noop {} {\bibfield  {journal} {\bibinfo  {journal} {Phys. Rev. Lett., \textbf{108}, 268303}\ } (\bibinfo {year} {2012})}\BibitemShut {NoStop}%
\bibitem [{\citenamefont {Ro}\ \emph {et~al.}(2021)\citenamefont {Ro}, \citenamefont {Kafri}, \citenamefont {Kardar},\ and\ \citenamefont {Tailleur}}]{densityfield}%
  \BibitemOpen
  \bibfield  {author} {\bibinfo {author} {\bibfnamefont {S.}~\bibnamefont {Ro}}, \bibinfo {author} {\bibfnamefont {Y.}~\bibnamefont {Kafri}}, \bibinfo {author} {\bibfnamefont {M.}~\bibnamefont {Kardar}}, \ and\ \bibinfo {author} {\bibfnamefont {J.}~\bibnamefont {Tailleur}},\ }\href@noop {} {\bibfield  {journal} {\bibinfo  {journal} {Phys. Rev. Lett. \textbf{126}, 048003}\ } (\bibinfo {year} {2021})}\BibitemShut {NoStop}%
\bibitem [{\citenamefont {Dauchot}\ and\ \citenamefont {Löwen}(2019)}]{chemicalfield}%
  \BibitemOpen
  \bibfield  {author} {\bibinfo {author} {\bibfnamefont {O.}~\bibnamefont {Dauchot}}\ and\ \bibinfo {author} {\bibfnamefont {H.}~\bibnamefont {Löwen}},\ }\href@noop {} {\bibfield  {journal} {\bibinfo  {journal} {J. Chem. Phys. \textbf{151}, 114901}\ } (\bibinfo {year} {2019})}\BibitemShut {NoStop}%
\bibitem [{\citenamefont {Toner}\ \emph {et~al.}(2018)\citenamefont {Toner}, \citenamefont {Guttenberg},\ and\ \citenamefont {Tu}}]{polfield}%
  \BibitemOpen
  \bibfield  {author} {\bibinfo {author} {\bibfnamefont {J.}~\bibnamefont {Toner}}, \bibinfo {author} {\bibfnamefont {N.}~\bibnamefont {Guttenberg}}, \ and\ \bibinfo {author} {\bibfnamefont {Y.}~\bibnamefont {Tu}},\ }\href@noop {} {\bibfield  {journal} {\bibinfo  {journal} {Phys. Rev. E \textbf{98}, 062604}\ } (\bibinfo {year} {2018})}\BibitemShut {NoStop}%
\bibitem [{\citenamefont {Das}\ \emph {et~al.}(2018)\citenamefont {Das}, \citenamefont {Kumar},\ and\ \citenamefont {Mishra}}]{pol2}%
  \BibitemOpen
  \bibfield  {author} {\bibinfo {author} {\bibfnamefont {R.}~\bibnamefont {Das}}, \bibinfo {author} {\bibfnamefont {M.}~\bibnamefont {Kumar}}, \ and\ \bibinfo {author} {\bibfnamefont {S.}~\bibnamefont {Mishra}},\ }\href@noop {} {\bibfield  {journal} {\bibinfo  {journal} {Phys. Rev. E \textbf{98}, 060602(R) (2018)}\ } (\bibinfo {year} {2018})}\BibitemShut {NoStop}%
\bibitem [{\citenamefont {Peruani}\ and\ \citenamefont {Aranson}(2018)}]{coldperuani}%
  \BibitemOpen
  \bibfield  {author} {\bibinfo {author} {\bibfnamefont {F.}~\bibnamefont {Peruani}}\ and\ \bibinfo {author} {\bibfnamefont {I.~S.}\ \bibnamefont {Aranson}},\ }\href@noop {} {\bibfield  {journal} {\bibinfo  {journal} {Phys. Rev. Lett. \textbf{120}, 238101}\ } (\bibinfo {year} {2018})}\BibitemShut {NoStop}%
\bibitem [{\citenamefont {Semwal}\ \emph {et~al.}(2021)\citenamefont {Semwal}, \citenamefont {Dikshit}, ,\ and\ \citenamefont {Mishra}}]{vivekepje}%
  \BibitemOpen
  \bibfield  {author} {\bibinfo {author} {\bibfnamefont {V.}~\bibnamefont {Semwal}}, \bibinfo {author} {\bibfnamefont {S.}~\bibnamefont {Dikshit}}, , \ and\ \bibinfo {author} {\bibfnamefont {S.}~\bibnamefont {Mishra}},\ }\href@noop {} {\bibfield  {journal} {\bibinfo  {journal} {Eur. Phys. J. E \textbf{44}, 20}\ } (\bibinfo {year} {2021})}\BibitemShut {NoStop}%
\bibitem [{\citenamefont {Volpe}\ \emph {et~al.}(2011)\citenamefont {Volpe}, \citenamefont {Buttinoni}, \citenamefont {Vogt}, \citenamefont {Kümmerer},\ and\ \citenamefont {Bechinger}}]{patternedBech}%
  \BibitemOpen
  \bibfield  {author} {\bibinfo {author} {\bibfnamefont {G.}~\bibnamefont {Volpe}}, \bibinfo {author} {\bibfnamefont {I.}~\bibnamefont {Buttinoni}}, \bibinfo {author} {\bibfnamefont {D.}~\bibnamefont {Vogt}}, \bibinfo {author} {\bibfnamefont {H.-J.}\ \bibnamefont {Kümmerer}}, \ and\ \bibinfo {author} {\bibfnamefont {C.}~\bibnamefont {Bechinger}},\ }\href@noop {} {\bibfield  {journal} {\bibinfo  {journal} {Soft Matter, \textbf{7}, 8810}\ } (\bibinfo {year} {2011})}\BibitemShut {NoStop}%
\bibitem [{\citenamefont {Qi}\ and\ \citenamefont {Dijkstra}(2015)}]{quench}%
  \BibitemOpen
  \bibfield  {author} {\bibinfo {author} {\bibfnamefont {W.}~\bibnamefont {Qi}}\ and\ \bibinfo {author} {\bibfnamefont {M.}~\bibnamefont {Dijkstra}},\ }\href@noop {} {\bibfield  {journal} {\bibinfo  {journal} {Soft Matter, \textbf{11}, 2852}\ } (\bibinfo {year} {2015})}\BibitemShut {NoStop}%
\bibitem [{\citenamefont {Liebchen}\ and\ \citenamefont {Löwen}(2018)}]{acc1}%
  \BibitemOpen
  \bibfield  {author} {\bibinfo {author} {\bibfnamefont {B.}~\bibnamefont {Liebchen}}\ and\ \bibinfo {author} {\bibfnamefont {H.}~\bibnamefont {Löwen}},\ }\href@noop {} {\bibfield  {journal} {\bibinfo  {journal} {Acc. Chem. Res. \textbf{51}, 2982}\ } (\bibinfo {year} {2018})}\BibitemShut {NoStop}%
\bibitem [{\citenamefont {Stark}(2018)}]{acc2}%
  \BibitemOpen
  \bibfield  {author} {\bibinfo {author} {\bibfnamefont {H.}~\bibnamefont {Stark}},\ }\href@noop {} {\bibfield  {journal} {\bibinfo  {journal} {Acc. Chem. Res. \textbf{51}, 2681}\ } (\bibinfo {year} {2018})}\BibitemShut {NoStop}%
\bibitem [{\citenamefont {Liebchen}\ \emph {et~al.}(2016)\citenamefont {Liebchen}, \citenamefont {Cates},\ and\ \citenamefont {Marenduzzo}}]{activerotors}%
  \BibitemOpen
  \bibfield  {author} {\bibinfo {author} {\bibfnamefont {B.}~\bibnamefont {Liebchen}}, \bibinfo {author} {\bibfnamefont {M.~E.}\ \bibnamefont {Cates}}, \ and\ \bibinfo {author} {\bibfnamefont {D.}~\bibnamefont {Marenduzzo}},\ }\href@noop {} {\bibfield  {journal} {\bibinfo  {journal} {Soft Matter, \textbf{12}, 7259}\ } (\bibinfo {year} {2016})}\BibitemShut {NoStop}%
\bibitem [{\citenamefont {Zhao}\ \emph {et~al.}(2023)\citenamefont {Zhao}, \citenamefont {Košmrlj},\ and\ \citenamefont {Datta}}]{arXiv}%
  \BibitemOpen
  \bibfield  {author} {\bibinfo {author} {\bibfnamefont {H.}~\bibnamefont {Zhao}}, \bibinfo {author} {\bibfnamefont {A.}~\bibnamefont {Košmrlj}}, \ and\ \bibinfo {author} {\bibfnamefont {S.~S.}\ \bibnamefont {Datta}},\ }\href@noop {} {\bibfield  {journal} {\bibinfo  {journal} {Phys. Rev. Lett. \textbf{131}, 118301}\ } (\bibinfo {year} {2023})}\BibitemShut {NoStop}%
\bibitem [{\citenamefont {Nejada}\ and\ \citenamefont {Najafi}(2019)}]{instability}%
  \BibitemOpen
  \bibfield  {author} {\bibinfo {author} {\bibfnamefont {M.~R.}\ \bibnamefont {Nejada}}\ and\ \bibinfo {author} {\bibfnamefont {A.}~\bibnamefont {Najafi}},\ }\href@noop {} {\bibfield  {journal} {\bibinfo  {journal} {Soft Matter, \textbf{15}, 3248}\ } (\bibinfo {year} {2019})}\BibitemShut {NoStop}%
\bibitem [{\citenamefont {Fadda}\ \emph {et~al.}(2023)\citenamefont {Fadda}, \citenamefont {Matoz-Fernandez}, \citenamefont {van Roij},\ and\ \citenamefont {Jabbari-Farouji}}]{softmatter2023}%
  \BibitemOpen
  \bibfield  {author} {\bibinfo {author} {\bibfnamefont {F.}~\bibnamefont {Fadda}}, \bibinfo {author} {\bibfnamefont {D.~A.}\ \bibnamefont {Matoz-Fernandez}}, \bibinfo {author} {\bibfnamefont {R.}~\bibnamefont {van Roij}}, \ and\ \bibinfo {author} {\bibfnamefont {S.}~\bibnamefont {Jabbari-Farouji}},\ }\href@noop {} {\bibfield  {journal} {\bibinfo  {journal} {Soft Matter, \textbf{10}, 1039}\ } (\bibinfo {year} {2023})}\BibitemShut {NoStop}%
\bibitem [{\citenamefont {Hu}\ \emph {et~al.}(2015)\citenamefont {Hu}, \citenamefont {Wysocki}, \citenamefont {Winkler},\ and\ \citenamefont {Gompper}}]{bact_morph}%
  \BibitemOpen
  \bibfield  {author} {\bibinfo {author} {\bibfnamefont {J.}~\bibnamefont {Hu}}, \bibinfo {author} {\bibfnamefont {A.}~\bibnamefont {Wysocki}}, \bibinfo {author} {\bibfnamefont {R.}~\bibnamefont {Winkler}}, \ and\ \bibinfo {author} {\bibfnamefont {G.}~\bibnamefont {Gompper}},\ }\href@noop {} {\bibfield  {journal} {\bibinfo  {journal} {Sci Rep \textbf{5}, 9586}\ } (\bibinfo {year} {2015})}\BibitemShut {NoStop}%
\bibitem [{\citenamefont {Pelton}\ \emph {et~al.}(2004)\citenamefont {Pelton}, \citenamefont {Ladavac},\ and\ \citenamefont {Grier}}]{morphology}%
  \BibitemOpen
  \bibfield  {author} {\bibinfo {author} {\bibfnamefont {M.}~\bibnamefont {Pelton}}, \bibinfo {author} {\bibfnamefont {K.}~\bibnamefont {Ladavac}}, \ and\ \bibinfo {author} {\bibfnamefont {D.}~\bibnamefont {Grier}},\ }\href@noop {} {\bibfield  {journal} {\bibinfo  {journal} {Phys. Rev. E \textbf{70}, 031108}\ } (\bibinfo {year} {2004})}\BibitemShut {NoStop}%
\bibitem [{\citenamefont {Choudhury}\ \emph {et~al.}(2017)\citenamefont {Choudhury}, \citenamefont {Straube}, \citenamefont {Fischer}, \citenamefont {Gibbs},\ and\ \citenamefont {Hoﬂing}}]{periodicsurface}%
  \BibitemOpen
  \bibfield  {author} {\bibinfo {author} {\bibfnamefont {U.}~\bibnamefont {Choudhury}}, \bibinfo {author} {\bibfnamefont {A.}~\bibnamefont {Straube}}, \bibinfo {author} {\bibfnamefont {P.}~\bibnamefont {Fischer}}, \bibinfo {author} {\bibfnamefont {J.}~\bibnamefont {Gibbs}}, \ and\ \bibinfo {author} {\bibfnamefont {F.}~\bibnamefont {Hoﬂing}},\ }\href@noop {} {\bibfield  {journal} {\bibinfo  {journal} {New J. Phys. \textbf{19}, 125010}\ } (\bibinfo {year} {2017})}\BibitemShut {NoStop}%
\bibitem [{\citenamefont {Pattanayak}\ \emph {et~al.}(2019)\citenamefont {Pattanayak}, \citenamefont {Das}, \citenamefont {Kumar},\ and\ \citenamefont {Mishra}}]{sudipta}%
  \BibitemOpen
  \bibfield  {author} {\bibinfo {author} {\bibfnamefont {S.}~\bibnamefont {Pattanayak}}, \bibinfo {author} {\bibfnamefont {R.}~\bibnamefont {Das}}, \bibinfo {author} {\bibfnamefont {M.}~\bibnamefont {Kumar}}, \ and\ \bibinfo {author} {\bibfnamefont {S.}~\bibnamefont {Mishra}},\ }\href@noop {} {\bibfield  {journal} {\bibinfo  {journal} {Eur. Phys. J. E \textbf{42}, 62}\ } (\bibinfo {year} {2019})}\BibitemShut {NoStop}%
\bibitem [{\citenamefont {Dolai}\ \emph {et~al.}(2018)\citenamefont {Dolai}, \citenamefont {Simha},\ and\ \citenamefont {Mishra}}]{softmatterpritha2018}%
  \BibitemOpen
  \bibfield  {author} {\bibinfo {author} {\bibfnamefont {P.}~\bibnamefont {Dolai}}, \bibinfo {author} {\bibfnamefont {A.}~\bibnamefont {Simha}}, \ and\ \bibinfo {author} {\bibfnamefont {S.}~\bibnamefont {Mishra}},\ }\href@noop {} {\bibfield  {journal} {\bibinfo  {journal} {Soft Matter, \textbf{14}, 6137}\ } (\bibinfo {year} {2018})}\BibitemShut {NoStop}%
\bibitem [{\citenamefont {Fily}\ and\ \citenamefont {Marchetti}(2012)}]{prl2012filly}%
  \BibitemOpen
  \bibfield  {author} {\bibinfo {author} {\bibfnamefont {Y.}~\bibnamefont {Fily}}\ and\ \bibinfo {author} {\bibfnamefont {M.~C.}\ \bibnamefont {Marchetti}},\ }\href@noop {} {\bibfield  {journal} {\bibinfo  {journal} {Phys. Rev. Lett. \textbf{108}, 235702}\ } (\bibinfo {year} {2012})}\BibitemShut {NoStop}%
\bibitem [{\citenamefont {Keller}\ and\ \citenamefont {Segel}(1970)}]{KS1}%
  \BibitemOpen
  \bibfield  {author} {\bibinfo {author} {\bibfnamefont {E.~F.}\ \bibnamefont {Keller}}\ and\ \bibinfo {author} {\bibfnamefont {L.~A.}\ \bibnamefont {Segel}},\ }\href@noop {} {\bibfield  {journal} {\bibinfo  {journal} {J. Theor. Biol., \textbf{26}, 399}\ } (\bibinfo {year} {1970})}\BibitemShut {NoStop}%
\bibitem [{\citenamefont {Keller}\ and\ \citenamefont {Segel}(1971)}]{KS2}%
  \BibitemOpen
  \bibfield  {author} {\bibinfo {author} {\bibfnamefont {E.~F.}\ \bibnamefont {Keller}}\ and\ \bibinfo {author} {\bibfnamefont {L.~A.}\ \bibnamefont {Segel}},\ }\href@noop {} {\bibfield  {journal} {\bibinfo  {journal} {J. Theor. Biol., \textbf{30}, 225}\ } (\bibinfo {year} {1971})}\BibitemShut {NoStop}%
\bibitem [{\citenamefont {Masoud}\ and\ \citenamefont {Shelley}(2014)}]{chemact1}%
  \BibitemOpen
  \bibfield  {author} {\bibinfo {author} {\bibfnamefont {H.}~\bibnamefont {Masoud}}\ and\ \bibinfo {author} {\bibfnamefont {M.~J.}\ \bibnamefont {Shelley}},\ }\href@noop {} {\bibfield  {journal} {\bibinfo  {journal} {Phys. Rev. Lett., \textbf{112}, 128304}\ } (\bibinfo {year} {2014})}\BibitemShut {NoStop}%
\bibitem [{\citenamefont {Nasouri}\ and\ \citenamefont {Golestanian}(2020)}]{chemact2}%
  \BibitemOpen
  \bibfield  {author} {\bibinfo {author} {\bibfnamefont {B.}~\bibnamefont {Nasouri}}\ and\ \bibinfo {author} {\bibfnamefont {R.}~\bibnamefont {Golestanian}},\ }\href@noop {} {\bibfield  {journal} {\bibinfo  {journal} {Phys. Rev. Lett., \textbf{124}, 168003}\ } (\bibinfo {year} {2020})}\BibitemShut {NoStop}%
\bibitem [{\citenamefont {Bialké}\ \emph {et~al.}(2012)\citenamefont {Bialké}, \citenamefont {Speck},\ and\ \citenamefont {Löwen}}]{jamming}%
  \BibitemOpen
  \bibfield  {author} {\bibinfo {author} {\bibfnamefont {J.}~\bibnamefont {Bialké}}, \bibinfo {author} {\bibfnamefont {T.}~\bibnamefont {Speck}}, \ and\ \bibinfo {author} {\bibfnamefont {H.}~\bibnamefont {Löwen}},\ }\href@noop {} {\bibfield  {journal} {\bibinfo  {journal} {Phys. Rev. Lett. \textbf{108}, 168301}\ } (\bibinfo {year} {2012})}\BibitemShut {NoStop}%
\bibitem [{\citenamefont {Buttinoni}\ \emph {et~al.}(2013)\citenamefont {Buttinoni}, \citenamefont {Bialké}, \citenamefont {Kümmel}, \citenamefont {Löwen}, \citenamefont {Bechinger},\ and\ \citenamefont {Speck}}]{prl2013exp}%
  \BibitemOpen
  \bibfield  {author} {\bibinfo {author} {\bibfnamefont {I.}~\bibnamefont {Buttinoni}}, \bibinfo {author} {\bibfnamefont {J.}~\bibnamefont {Bialké}}, \bibinfo {author} {\bibfnamefont {F.}~\bibnamefont {Kümmel}}, \bibinfo {author} {\bibfnamefont {H.}~\bibnamefont {Löwen}}, \bibinfo {author} {\bibfnamefont {C.}~\bibnamefont {Bechinger}}, \ and\ \bibinfo {author} {\bibfnamefont {T.}~\bibnamefont {Speck}},\ }\href@noop {} {\bibfield  {journal} {\bibinfo  {journal} {Phys. Rev. Lett. \textbf{110}, 238301}\ } (\bibinfo {year} {2013})}\BibitemShut {NoStop}%
\bibitem [{\citenamefont {Redner}\ \emph {et~al.}(2013)\citenamefont {Redner}, \citenamefont {Hagan},\ and\ \citenamefont {Baskaran}}]{peclet}%
  \BibitemOpen
  \bibfield  {author} {\bibinfo {author} {\bibfnamefont {G.~S.}\ \bibnamefont {Redner}}, \bibinfo {author} {\bibfnamefont {M.~F.}\ \bibnamefont {Hagan}}, \ and\ \bibinfo {author} {\bibfnamefont {A.}~\bibnamefont {Baskaran}},\ }\href@noop {} {\bibfield  {journal} {\bibinfo  {journal} {Phys. Rev. Lett. \textbf{110}, 055701}\ } (\bibinfo {year} {2013})}\BibitemShut {NoStop}%
\bibitem [{\citenamefont {Baskaran}\ and\ \citenamefont {Marchetti}(2008)}]{beta1}%
  \BibitemOpen
  \bibfield  {author} {\bibinfo {author} {\bibfnamefont {A.}~\bibnamefont {Baskaran}}\ and\ \bibinfo {author} {\bibfnamefont {M.~C.}\ \bibnamefont {Marchetti}},\ }\href@noop {} {\bibfield  {journal} {\bibinfo  {journal} {Phys. Rev. Lett. \textbf{101}, 268101}\ } (\bibinfo {year} {2008})}\BibitemShut {NoStop}%
\bibitem [{\citenamefont {Howse}\ \emph {et~al.}(2007)\citenamefont {Howse}, \citenamefont {Jones}, \citenamefont {Ryan}, \citenamefont {Gough}, \citenamefont {Vafabakhsh},\ and\ \citenamefont {Golestanian}}]{beta2}%
  \BibitemOpen
  \bibfield  {author} {\bibinfo {author} {\bibfnamefont {J.~R.}\ \bibnamefont {Howse}}, \bibinfo {author} {\bibfnamefont {R.~A.~L.}\ \bibnamefont {Jones}}, \bibinfo {author} {\bibfnamefont {A.~J.}\ \bibnamefont {Ryan}}, \bibinfo {author} {\bibfnamefont {T.}~\bibnamefont {Gough}}, \bibinfo {author} {\bibfnamefont {R.}~\bibnamefont {Vafabakhsh}}, \ and\ \bibinfo {author} {\bibfnamefont {R.}~\bibnamefont {Golestanian}},\ }\href@noop {} {\bibfield  {journal} {\bibinfo  {journal} {Phys. Rev. Lett. \textbf{99}, 048102}\ } (\bibinfo {year} {2007})}\BibitemShut {NoStop}%
\bibitem [{\citenamefont {Mertens}\ and\ \citenamefont {Moore}(2012)}]{percolation}%
  \BibitemOpen
  \bibfield  {author} {\bibinfo {author} {\bibfnamefont {S.}~\bibnamefont {Mertens}}\ and\ \bibinfo {author} {\bibfnamefont {C.}~\bibnamefont {Moore}},\ }\href@noop {} {\bibfield  {journal} {\bibinfo  {journal} {Phys. Rev. E \textbf{86}, 061109}\ } (\bibinfo {year} {2012})}\BibitemShut {NoStop}%
\bibitem [{\citenamefont {Smit}\ and\ \citenamefont {Frenkel}(1991)}]{equilibiurmliquidvapour}%
  \BibitemOpen
  \bibfield  {author} {\bibinfo {author} {\bibfnamefont {B.}~\bibnamefont {Smit}}\ and\ \bibinfo {author} {\bibfnamefont {D.}~\bibnamefont {Frenkel}},\ }\href@noop {} {\bibfield  {journal} {\bibinfo  {journal} {J. Chem. Phys. \textbf{94}, 5663–5668}\ } (\bibinfo {year} {1991})}\BibitemShut {NoStop}%
\bibitem [{\citenamefont {Mijalkov}\ \emph {et~al.}(2016)\citenamefont {Mijalkov}, \citenamefont {McDaniel}, \citenamefont {Wehr},\ and\ \citenamefont {Volpe}}]{photo}%
  \BibitemOpen
  \bibfield  {author} {\bibinfo {author} {\bibfnamefont {M.}~\bibnamefont {Mijalkov}}, \bibinfo {author} {\bibfnamefont {A.}~\bibnamefont {McDaniel}}, \bibinfo {author} {\bibfnamefont {J.}~\bibnamefont {Wehr}}, \ and\ \bibinfo {author} {\bibfnamefont {G.}~\bibnamefont {Volpe}},\ }\href@noop {} {\bibfield  {journal} {\bibinfo  {journal} {Phys. Rev. X \textbf{6}, 011008}\ } (\bibinfo {year} {2016})}\BibitemShut {NoStop}%
\bibitem [{\citenamefont {Liebchen}\ \emph {et~al.}(2018)\citenamefont {Liebchen}, \citenamefont {Monderkamp}, \citenamefont {Hagen},\ and\ \citenamefont {Löwen}}]{visco}%
  \BibitemOpen
  \bibfield  {author} {\bibinfo {author} {\bibfnamefont {B.}~\bibnamefont {Liebchen}}, \bibinfo {author} {\bibfnamefont {P.}~\bibnamefont {Monderkamp}}, \bibinfo {author} {\bibfnamefont {B.~T.}\ \bibnamefont {Hagen}}, \ and\ \bibinfo {author} {\bibfnamefont {H.}~\bibnamefont {Löwen}},\ }\href@noop {} {\bibfield  {journal} {\bibinfo  {journal} {Phys. Rev. Lett. \textbf{120}, 208002}\ } (\bibinfo {year} {2018})}\BibitemShut {NoStop}%
\bibitem [{\citenamefont {Cohen}\ \emph {et~al.}(2014)\citenamefont {Cohen}, \citenamefont {Nelson}, ,\ and\ \citenamefont {Maharbiz}}]{electro}%
  \BibitemOpen
  \bibfield  {author} {\bibinfo {author} {\bibfnamefont {D.~J.}\ \bibnamefont {Cohen}}, \bibinfo {author} {\bibfnamefont {W.~J.}\ \bibnamefont {Nelson}}, , \ and\ \bibinfo {author} {\bibfnamefont {M.~M.}\ \bibnamefont {Maharbiz}},\ }\href@noop {} {\bibfield  {journal} {\bibinfo  {journal} {Nat. Mater. \textbf{13}, 409}\ } (\bibinfo {year} {2014})}\BibitemShut {NoStop}%
\bibitem [{\citenamefont {Golestanian}(2012)}]{thermo}%
  \BibitemOpen
  \bibfield  {author} {\bibinfo {author} {\bibfnamefont {R.}~\bibnamefont {Golestanian}},\ }\href@noop {} {\bibfield  {journal} {\bibinfo  {journal} {Phys. Rev. Lett. \textbf{108}, 038303}\ } (\bibinfo {year} {2012})}\BibitemShut {NoStop}%
\bibitem [{\citenamefont {Bäuerle}\ \emph {et~al.}(2020)\citenamefont {Bäuerle}, \citenamefont {Löffler},\ and\ \citenamefont {Bechinger}}]{swarm}%
  \BibitemOpen
  \bibfield  {author} {\bibinfo {author} {\bibfnamefont {T.}~\bibnamefont {Bäuerle}}, \bibinfo {author} {\bibfnamefont {R.~C.}\ \bibnamefont {Löffler}}, \ and\ \bibinfo {author} {\bibfnamefont {C.}~\bibnamefont {Bechinger}},\ }\href@noop {} {\bibfield  {journal} {\bibinfo  {journal} {Nat Commun \textbf{11}, 2547}\ } (\bibinfo {year} {2020})}\BibitemShut {NoStop}%
\bibitem [{\citenamefont {J.~M.~Peters}(2022)}]{honeybee}%
  \BibitemOpen
  \bibfield  {author} {\bibinfo {author} {\bibfnamefont {L.~M.}\ \bibnamefont {J.~M.~Peters}, \bibfnamefont {O.~Peleg~and}},\ }\href@noop {} {\bibfield  {journal} {\bibinfo  {journal} {J Exp Biol \textbf{225} (5):jeb242234}\ } (\bibinfo {year} {2022})}\BibitemShut {NoStop}%
\bibitem [{\citenamefont {Casiulis}\ and\ \citenamefont {Levine}(2022)}]{swarming}%
  \BibitemOpen
  \bibfield  {author} {\bibinfo {author} {\bibfnamefont {M.}~\bibnamefont {Casiulis}}\ and\ \bibinfo {author} {\bibfnamefont {D.}~\bibnamefont {Levine}},\ }\href@noop {} {\bibfield  {journal} {\bibinfo  {journal} {Phys. Rev. E \textbf{106}, 044611}\ } (\bibinfo {year} {2022})}\BibitemShut {NoStop}%
\bibitem [{\citenamefont {Tung}\ \emph {et~al.}(2016)\citenamefont {Tung}, \citenamefont {Harder}, \citenamefont {Valeriani},\ and\ \citenamefont {Cacciuto}}]{paper2}%
  \BibitemOpen
  \bibfield  {author} {\bibinfo {author} {\bibfnamefont {C.}~\bibnamefont {Tung}}, \bibinfo {author} {\bibfnamefont {J.}~\bibnamefont {Harder}}, \bibinfo {author} {\bibfnamefont {C.}~\bibnamefont {Valeriani}}, \ and\ \bibinfo {author} {\bibfnamefont {A.}~\bibnamefont {Cacciuto}},\ }\href@noop {} {\bibfield  {journal} {\bibinfo  {journal} {Soft Matter \textbf{12}, 555-561}\ } (\bibinfo {year} {2016})}\BibitemShut {NoStop}%
\bibitem [{\citenamefont {Vahabli}\ and\ \citenamefont {Vicsek}(2023)}]{vicseknatphys}%
  \BibitemOpen
  \bibfield  {author} {\bibinfo {author} {\bibfnamefont {D.}~\bibnamefont {Vahabli}}\ and\ \bibinfo {author} {\bibfnamefont {T.}~\bibnamefont {Vicsek}},\ }\href@noop {} {\bibfield  {journal} {\bibinfo  {journal} {Communications Physics \textbf{6}, 56}\ } (\bibinfo {year} {2023})}\BibitemShut {NoStop}%
\bibitem [{\citenamefont {Toner}\ and\ \citenamefont {Tu}(1998)}]{tonertupre1998}%
  \BibitemOpen
  \bibfield  {author} {\bibinfo {author} {\bibfnamefont {J.}~\bibnamefont {Toner}}\ and\ \bibinfo {author} {\bibfnamefont {Y.}~\bibnamefont {Tu}},\ }\href@noop {} {\bibfield  {journal} {\bibinfo  {journal} {Phys. Rev. E \textbf{58}, 4828}\ } (\bibinfo {year} {1998})}\BibitemShut {NoStop}%
\bibitem [{\citenamefont {Stenhammar}\ \emph {et~al.}(2013)\citenamefont {Stenhammar}, \citenamefont {Tiribocchi}, \citenamefont {Allen}, \citenamefont {Marenduzzo},\ and\ \citenamefont {Cates}}]{catesprl2013}%
  \BibitemOpen
  \bibfield  {author} {\bibinfo {author} {\bibfnamefont {J.}~\bibnamefont {Stenhammar}}, \bibinfo {author} {\bibfnamefont {A.}~\bibnamefont {Tiribocchi}}, \bibinfo {author} {\bibfnamefont {R.~J.}\ \bibnamefont {Allen}}, \bibinfo {author} {\bibfnamefont {D.}~\bibnamefont {Marenduzzo}}, \ and\ \bibinfo {author} {\bibfnamefont {M.~E.}\ \bibnamefont {Cates}},\ }\href@noop {} {\bibfield  {journal} {\bibinfo  {journal} {Phys. Rev. Lett. \textbf{111}, 145702}\ } (\bibinfo {year} {2013})}\BibitemShut {NoStop}%
\bibitem [{\citenamefont {Liebchen}\ \emph {et~al.}(2015)\citenamefont {Liebchen}, \citenamefont {Marenduzzo}, \citenamefont {Pagonabarraga},\ and\ \citenamefont {Cates}}]{paper1}%
  \BibitemOpen
  \bibfield  {author} {\bibinfo {author} {\bibfnamefont {B.}~\bibnamefont {Liebchen}}, \bibinfo {author} {\bibfnamefont {D.}~\bibnamefont {Marenduzzo}}, \bibinfo {author} {\bibfnamefont {I.}~\bibnamefont {Pagonabarraga}}, \ and\ \bibinfo {author} {\bibfnamefont {M.~E.}\ \bibnamefont {Cates}},\ }\href@noop {} {\bibfield  {journal} {\bibinfo  {journal} {Phys. Rev. Lett. \textbf{115}, 258301}\ } (\bibinfo {year} {2015})}\BibitemShut {NoStop}%
\bibitem [{\citenamefont {Klamser}\ \emph {et~al.}(2021)\citenamefont {Klamser}, \citenamefont {Dauchot},\ and\ \citenamefont {Tailleur}}]{KMC}%
  \BibitemOpen
  \bibfield  {author} {\bibinfo {author} {\bibfnamefont {J.~U.}\ \bibnamefont {Klamser}}, \bibinfo {author} {\bibfnamefont {O.}~\bibnamefont {Dauchot}}, \ and\ \bibinfo {author} {\bibfnamefont {J.}~\bibnamefont {Tailleur}},\ }\href@noop {} {\bibfield  {journal} {\bibinfo  {journal} {Phys. Rev. Lett. \textbf{127}, 150602}\ } (\bibinfo {year} {2021})}\BibitemShut {NoStop}%
\bibitem [{\citenamefont {Soddemann}\ \emph {et~al.}(2003)\citenamefont {Soddemann}, \citenamefont {Dünweg},\ and\ \citenamefont {Kremer}}]{NEMD}%
  \BibitemOpen
  \bibfield  {author} {\bibinfo {author} {\bibfnamefont {T.}~\bibnamefont {Soddemann}}, \bibinfo {author} {\bibfnamefont {B.}~\bibnamefont {Dünweg}}, \ and\ \bibinfo {author} {\bibfnamefont {K.}~\bibnamefont {Kremer}},\ }\href@noop {} {\bibfield  {journal} {\bibinfo  {journal} {Phys. Rev. E \textbf{68}, 046702}\ } (\bibinfo {year} {2003})}\BibitemShut {NoStop}%
\end{thebibliography}%

\end{document}